\documentclass[prl,twocolumn,floats,showpacs,superscriptaddress]{revtex4}
\usepackage{amssymb}
\usepackage{graphicx}
\usepackage{dcolumn}
\usepackage{amsmath}
\usepackage{bm}
\usepackage{epsfig}

\begin{document}

\title{$k$-core organization of complex networks}
\author{S. N. Dorogovtsev}
\affiliation{Departamento de F{\'\i}sica da Universidade de Aveiro, 3810-193 Aveiro,
Portugal}
\affiliation{Ioffe Physico-Technical Institute, 194021 St. Petersburg, Russia}
\author{A. V. Goltsev}
\affiliation{Departamento de F{\'\i}sica da Universidade de Aveiro, 3810-193 Aveiro,
Portugal}
\affiliation{Ioffe Physico-Technical Institute, 194021 St. Petersburg, Russia}
\author{J. F. F. Mendes}
\affiliation{Departamento de F{\'\i}sica da Universidade de Aveiro, 3810-193 Aveiro,
Portugal}

\begin{abstract} 

We analytically describe the architecture of randomly damaged uncorrelated networks as a set of successively enclosed substructures --- $k$-cores. The $k$-core is the largest subgraph where vertices have at least $k$ interconnections. 
We find the structure of $k$-cores, their sizes, and their birth points --- the bootstrap percolation thresholds.  
We show that in networks with a finite 
mean number $z_{2}$ of the second-nearest neighbors,  
the 
emergence of a $k$-core 
is a 
hybrid phase transition. 
In contrast, if $z_{2}$ diverges, the networks contain an infinite sequence of $k$-cores which are ultra-robust against random damage.
\end{abstract}

\maketitle





\emph{Introduction.}---Extracting and indexing highly interconnected parts
of complex networks---communities, cliques, cores, etc.---as well as finding
relations between these substructures is an issue of topical interest in
network research, see, e.g., Refs.~\cite{gm02,pdfv05}. This decomposition
helps one to describe the complex topologies of real-world networks. In this
respect, the notion of $k$-core is of fundamental importance \cite{b84,s83}. 
The $k$-core may be obtained in
the following way. Remove from a graph all vertices of degree less than $k$. 
Some of the rest vertices may remain with less than $%
k$ edges. Then remove these vertices, and so on until no further removal is
possible. The result, if it exists, is the $k${\em-core}. Thus, a network is
organized as a set of successively enclosed $k$-cores, similarly to a
Russian nesting doll.

The $k$-core decomposition was recently applied to a number of real-world
networks (the Internet, the WWW, cellular networks, etc.) \cite%
{aabv05,k05,wa05} and 
was 
turned out to be an important tool for visualization
of complex networks and interpretation of cooperative processes in them. 
Rich $k$-core architectures of real networks was revealed. 
Furthermore, a $k$-core related Jellyfish model \cite{tpsf01} is one of the 
popular models of the Autonomous System graph of the Internet. 
The
notion of the $k$-core is a natural generalization of the giant connected
component in the ordinary percolation 
\cite{ajb00,cah02,cnsw00} 
(for another possible generalization, see clique percolation in Ref. \cite{dpv05}). 
Impressively, the giant connected component of an infinite network with a
heavy-tailed degree distribution is robust against random damage of the net.
The $k$-core percolation implies the emergence of a giant $k$-core below a
threshold concentration of vertices or edges removed at random. In physics,
the $k$-core percolation (bootstrap percolation) on the Bethe lattice
was introduced in Ref. \cite{clr79} for describing some magnetic materials.
Note that the $k\geq 3$-core
percolation is an unusual, hybrid phase transition with a jump of the order
parameter as at a first order phase transition but also with strong critical
fluctuations as at a continuous phase transition \cite{clr79,slc04}. The $k$%
-core decomposition of a random graph was formulated as a mathematical
problem in Refs.~\cite{b84,s83}. This attracted much attention\thinspace of
mathematicians \cite{psw96,fr04}, but actually only the criteria of
emergence of $k$-cores in basic random networks were found.


\begin{figure}[b]
\epsfxsize=26mm 
\centerline{\epsffile{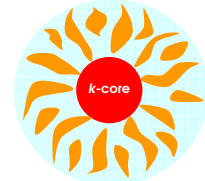}}
\caption{ The structure of a network with a $k$-core. The $k$-core is the
internal circle. Vertices of degrees smaller than $k$ form the light-grey
area. The dark-grey regions are numerous finite clusters with those vertices
of degrees $q \geq k$, which do not belong to the $k$-core. These clusters
are either connected to the $k$-core by less than $k$ edges or isolated from
it. }
\label{sun}
\end{figure}


In this Letter we derive exact equations describing the $k$-core
organization of a randomly damaged uncorrelated network with an arbitrary
degree distribution. This allows us to obtain the sizes and other structural
characteristics of $k$-cores in a variety of damaged and undamaged random
networks and find the nature of the $k$-core percolation in complex
networks. We apply our general results to the classical random graphs and to
scale-free networks, in particular, to empirical router-level Internet maps.
We find that not only the giant connected components in infinite networks
with slowly decreasing degree distributions are resilient against random
damage, as was known, but their entire $k$-core architectures are robust.

\emph{Basic equations.}---We consider an uncorrelated network---a maximally
random graph with a given degree distribution $P(q)$---the so-called
configuration model. We assume that a fraction $Q\equiv 1-p$ of the vertices
in this network are removed at random. 
The $k$-core extracting procedure
results in the structure of the network with a $k$-core depicted in 
Fig.~\ref{sun}. 

Taking into account the tree-like structure of the infinite sparse configuration model shows that the $k$-core coincides with the infinite $(k{-}1)$-ary subtree \cite{remark}. 
(The $m$-ary tree is a tree, where all vertices have branching at least $m$.) 
Let $R$ be the probability that a given end of an edge of a network 
is not the root of an infinite $(k{-}1)$-ary subtree. 
Then a vertex belongs to the $k$-core 
if at least $k$ its neighbors 
are roots of infinite $(k{-}1)$-ary subtrees. 
So the probability that a vertex is in the $k$-core is  
\begin{equation}
M(k)=p\sum\limits_{q\geqslant
k}^{{}}P(q)\sum\limits_{n=k}^{q}C_{n}^{q}R^{q-n}(1-R)^{n} ,   
\label{k-core}
\end{equation}
where $C_{n}^{m}=m!/(m-n)!n!$. 
Note that for the ordinary percolation
we must set $k=1$ in 
this equation. 

An end of an edge is not a root of an infinite $(k{-}1)$-ary subtree if at most $k{-}2$ its children branches are roots of infinite $(k{-}1)$-ary subtrees.  This leads to the following equation for $R$:   
\begin{equation}
\!R = 1{-}p{+}p\sum_{n=0}^{k-2} \left[\, \sum_{i=n}^\infty \frac{(i{+}1)P(i{+}1)}
{z_{1}}\, C_n^i R^{i-n} (1{-}R)^n \right] 
\!\!{.}\!\!\! 
\label{R}
\end{equation} 
Let us explain this equation. 
(i) The first term, $1{-}p\equiv Q$, 
is the probability that 
the end of the edge 
is unoccupied.  
(ii) $C_n^i R^{i-n} (1-R)^n$ is the probability that if a given end of the edge has $i$ children (i.e., other edges than the starting edge), then exactly $n$ of them are roots of infinite $(k{-}1)$-ary subtrees. 
$(i{+}1)P(i{+}1)/z_{1}$  is the probability that a randomly chosen edge leads to a
vertex with branching $i$. $z_{1}=\sum\nolimits_{q}qP(q)$ is the mean number of
the nearest neighbors of a vertex in the graph. 
Thus, in the square brackets, we present the probability that a given end of the edge has exactly $n$ edges, which are roots of infinite $(k-1)$-ary subtrees. 
(iii) 
Finally, 
we take into account that $n$ must be at most $k-2$. 

The sum $\sum_{n=0}^{k-2}$ in Eq.~(\ref{R}) may be 
rewritten as: 
\begin{equation}
\Phi _{k}(R)\!\! = \!\!\sum\limits_{n=0}^{k-2}\frac{(1-R)^{n}}{n!}\frac{d^{n}}{dR^{n}}G_{1}(R),
\label{F1}
\end{equation}
where $G_{1}(x)=z_{1}^{-1}\sum%
\nolimits_{q}P(q)qx^{q-1}=z_{1}^{-1}dG_{0}(x)/dx$, and $G_{0}(x)=\sum%
\nolimits_{q}P(q)x^{q}$ \cite{nsw01}. 
Then 
Eq.~(\ref{R}) takes the 
form: 
\begin{equation}
R=1-p+p\Phi _{k}(R).  
\label{R2}
\end{equation}
In the case $p=1$, Eq.~(\ref{R2}) was recently obtained in \cite{fr04}. If
Eq.~(\ref{R2}) has only the trivial solution $R\!=\!1$, there is no giant $k$%
-core. The emergence of a nontrivial solution corresponds to the birth of
the giant $k$-core. It is the lowest nontrivial solution $R\!<\!1$ that
describes the $k$-core.

Let us define a function 
\begin{equation}
f_{k}(R)=[1-\Phi _{k}(R)]/(1-R)  
.
\label{fk}
\end{equation}
This function is positive in the range $R\in \lbrack 0,1)$ and, in networks
with 
a finite mean number of the second neighbors of a
vertex, $z_{2}=\sum_{q}q(q-1)P(q)$, 
it tends to zero in the limit $R\rightarrow
1$ as $f_{k}(R)\propto (1-R)^{k-2}$. In terms of the function $f_{k}(R)$,
Eq. (\ref{R}) 
is especially simple: 
\begin{equation}
pf_{k}(R)=1.  \label{f}
\end{equation}
Depending on $P(q)$, with increasing $R$, $f_{k}(R)$ either (i)
monotonously decreases from $f_{k}(0)<1$ to $f_{k}(1)=0$, or (ii) at first
increases, then approaches a maximum at $R_{\max }\in (0,1)$, and finally
tends to zero at $R\rightarrow 1$. Therefore Eq. (\ref{f}) has a non-trivial
solution $R<1$ if 
\begin{equation}
p\max_{R\in \lbrack 0,1)}f_{k}(R)\geqslant 1.  \label{criterion}
\end{equation}
This is the criterion for the emergence of the giant $k$-core in a randomly
damaged uncorrelated network. The equality in Eq. (\ref{criterion}) takes
place at a critical concentration $p_{c}(k)$ when the line $y(R)=1/p_{c}(k)$
touches the maximum of $f_{k}(R)$. Therefore the threshold of the $k$-core
percolation is determined by two equations: 
\begin{equation}
p_{c}(k)=1/f_{k}(R_{\max }),\ \ \ \ \ \ 0=f_{k}^{\prime }(R_{\max }).
\label{cp1}
\end{equation}
$R_{\max }$ is the value of the order parameter at the birth point of the $k$%
-core. At $p<p_{c}(k)$ there is only the trivial solution $R=1$.

At $k=2$, Eq.~(\ref{R2}) describes the ordinary percolation in a random
uncorrelated graph \cite{cah02,cnsw00}. 
In this case, in infinite networks we have $R_{\max }\rightarrow 1$, and the
criterion (\ref{criterion}) is reduced to the standard condition for
existence of the giant connected component:  
$pG_{1}^{\prime }(1)=pz_{2}/z_{1}\geqslant 1$. 

Let us find 
$R$ near the $k\!\geq \!3$-core percolation transition in a network with a
finite $z_{2}$. We examine Eq. (\ref{R2}) for $R=R_{\max }+r$ and $%
p=p_{c}(k)+\epsilon $ with $\epsilon ,\left\vert r\right\vert \ll 1$. Note
that at $k\geqslant 3$, $\Phi _{k}(R)$ is an analytical function in the
range $R\in \lbrack 0,1)$. It means that the expansion of $\Phi _{k}(R+r)$
over $r$ contains no singular term at $R\in \lbrack 0,1)$. Substituting this
expansion into Eq. (\ref{R2}), in the leading order, we find 
\begin{equation}
R_{\max }-R\propto [p-p_{c}(k)]^{1/2},   
\label{expR}
\end{equation} 
i.e., the combination of a jump and the square root critical singularity. 
The origin of this singularity is an intriguing problem of the hybrid phase transition. 

The structure of the $k$-core is
essentially determined by its degree distribution which we find to be 
\begin{equation}
P_{k}(q)=\frac{p}{M(k)}\sum\limits_{q^{\prime }\geqslant q}P(q^{\prime
})C_{q}^{q^{\prime }}R^{q^{\prime }-q}(1-R)^{q}.  \label{z1(k)}
\end{equation}
The mean degree of the $k$-core vertices is $z_{1}(k)=\sum_{q\geq
k}P_{k}(q)q $. The $k$-core of a given graph contains the $k+1$-core as a
subgraph. Vertices which belong to the $k$-core, but do not belong to the $%
k+1$-core, form the $k$-shell of the relative size $S(k)=M(k)-M(k+1)$.

We apply our general results to two basic networks.


\begin{figure}[b]
\par
\begin{center}
\scalebox{0.29}{\includegraphics[angle=270]{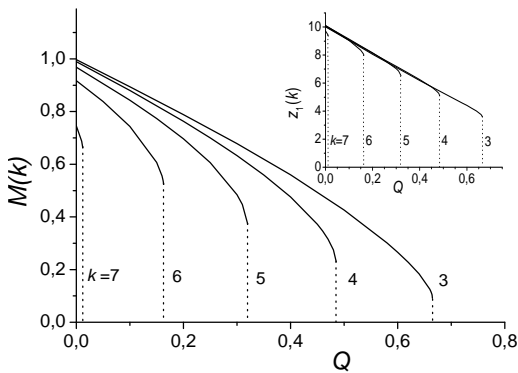}}
\end{center}
\caption{ The size of the $k$-core, $M(k)$, in the Erd\H{o}s-R\'{e}nyi
random graph with the mean degree $z_{1}=10$ versus the concentration $Q$ of
vertices removed at random. The highest core disappears at a very
low concentration $Q\approx 1.2\%$ in contrast to the ordinary percolation
threshold $Q\approx 90\%$. The inset shows the mean degree $z_{1}(k)$ of
vertices in the $k$-core. }
\label{dER}
\end{figure}


\emph{Erd\H{o}s-R\'{e}nyi (ER) graphs.}---These random graphs have the Poisson degree distribution $P(q)\!=\!z_{1}^{q}\exp(-z_{1})/q!$, where $z_{1}$ is the mean degree. 
In this case, $G_{0}(x)=G_{1}(x)=\exp [z_{1}(x\!-\!1)]$. In Eq.~(\ref{R2}), $%
\Phi _{k}(R)=\Gamma \lbrack k\!-\!1,z_{1}(1\!-\!R)]/\Gamma (k-1)$, where $%
\Gamma (n,x)$ is the incomplete gamma function. From Eq.~(\ref{k-core}) we
get the size of the $k$-core: 
\begin{equation}
M(k)=p\{1-\Gamma \lbrack k,z_{1}(1-R)]/\Gamma (k)\}
,  
\label{ER-core}
\end{equation}
where $R$ is the solution of Eq.~(\ref{R2}). The degree distribution in the $%
k$-core is 
$P_{k}(q{\geq }k)=pz_{1}^{q}(1{-}R)^{q} e^{-z_1(1{-}R)}/[M(k)q!]$. 
Our numerical
calculations revealed that at $p=1$, the highest $k$-core 
increases almost linearly with 
$z_{1}$, namely, $k_{h}\approx 0.78z_{1}$ 
at $z_{1}\lesssim 500$. Furthermore, 
the mean degree $z_{1}(k)$ in the $k$-core weakly depends on $k$: 
$z_{1}(k)\approx z_{1}$.

\begin{figure}[b]
\scalebox{0.34}{\includegraphics[angle=270]{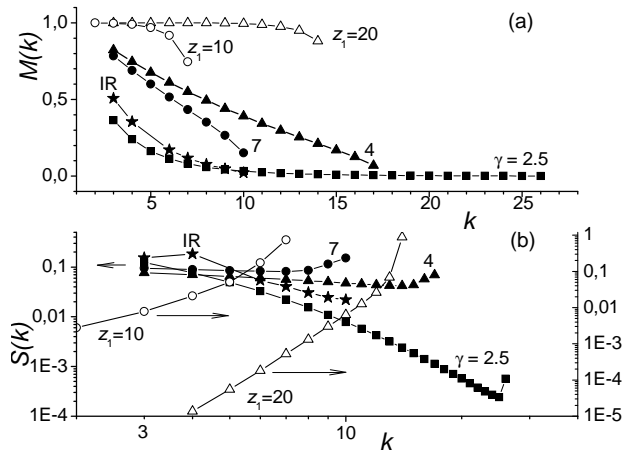}}
\caption{ The relative sizes of the $k$-cores, $M(k)$, panel (a), and $k$-shells, $S(k)$%
, panel (b), in the Erd\H os-R\'{e}nyi graphs with $z_1=10$ and $20$;
scale-free networks with $\protect\gamma =2.5,4$ and $7$, and an
uncorrelated network with the degree distribution of the router-level
Internet map (IR). The minimum degree in the scale-free networks is $q_0=1$.
In the case $\protect\gamma=2.5$, the maximum degree in the network is $q_{\text{cut}}= 2\,000$, and $c=2$; for $\protect\gamma=4$ and $7$, $c=30$ and $%
50$, respectively. }
\label{kCores}
\end{figure}

Fig.~\ref{dER} shows the dependence of the size of the $k$-cores, $M(k)$, on
the concentration $Q=1-p$ of the vertices removed at random. Note that
counterintuitively, it is the highest $k$-core---the central, most
interconnected part of a network---that is destroyed primarily. The inset of
Fig.~\ref{dER} shows that with increasing damage $Q$, the mean degree 
$z_{1}(k)$ decreases. The $k$-cores disappear consecutively, starting from the highest
core. The $k$-core structure of the undamaged ER 
graphs is displayed in Fig.~\ref{kCores}.

\emph{Scale-free networks.}---We consider uncorrelated networks with a
degree distribution $P(q)\propto (q+c)^{-\gamma }$. Let us start with the
case of $\gamma >3$, where $z_{2}$ is finite. It turns out that the
existence of $k$-cores is determined by the complete form of the degree
distribution including its low degree region. It was proved in Ref. \cite%
{fr04} that there is no $k\!\geqslant \!3$-core in a 
graph with the minimal degree $q_{0}=1$, $\gamma \geq 3$, and $c=0$. 
We find that the $k$-cores emerge as $c$ increases. The $k$-core structure
of scale-free graphs is represented in Fig. \ref{kCores}. The relative sizes
of the giant $k$-cores in the scale-free networks are smaller than in the ER 
graphs. As $z_{2}$ is finite, 
the 
$k\!\geq \!3$-core percolation 
at $\gamma \!>\!3$ is the 
hybrid phase transition%
. This is in contrast to the ordinary percolation in scale-free networks,
where behavior is non-standard if $\gamma \leq 4$ \cite{cah02}.


The case $2<\gamma \leqslant 3$ is realized in most important real-world
networks. With $\gamma $ in this range, $z_{2}$ diverges if $N\rightarrow
\infty $. In the leading order in $1-R\ll 1$, Eq.~(\ref{fk}) gives $%
f_{k}(R)\cong (q_{0}/k)^{\gamma -2}(1-R)^{-(3-\gamma )}$. From Eq. (\ref{f})
we find the order parameter $R$
. Substituting this solution into Eq. (\ref{k-core}), in the leading order
in $1-R$ we find that the size of the $k$-core decreases with increasing $k$%
: 
\begin{equation}
M(k)=p[q_{0}(1\!-\!R)/k]^{\gamma -1}=p^{2/(3-\gamma )}(q_{0}/k)^{(\gamma
-1)/(3-\gamma )}\!.  \label{k-core2}
\end{equation}

The divergence of $f_{k}(R)$ at $R\rightarrow 1$ means that the percolation
threshold $p_{c}(k)$ tends to zero as $N\rightarrow \infty $. The $k$-core
percolation transition in this limit is of infinite order similarly to the
ordinary percolation \cite{cah02}. As 
$k_{h}(N\!\rightarrow \!\infty )\rightarrow \infty $, there is an infinite
sequence of successively enclosed $k$-cores%
. One has to remove at random almost all vertices in order to destroy any of
these cores.

Eq.~(\ref{z1(k)}) allows us to find the degree distribution of $k$-cores in
scale-free networks. For $\gamma >2$ and $k\gg 1$, $P_{k}(q\gg k)\approx
(\gamma -1)k^{\gamma -1}q^{-\gamma }$. The mean degree $z_{1}(k)$ in the $k$%
-core grows linearly with $k$: $z_{1}(k)\approx kz_{1}/q_{0}$ in contrast to
the Erd\H{o}s-R\'{e}nyi graphs.

\emph{Finite-size effect.}---
The finiteness of the scale-free networks with $2<\gamma <3$ essentially
determines their $k$-core organization. We introduce a size dependent cutoff 
$q_{\text{cut}}(N)$ of the degree distribution. 
Here $q_{\text{cut}}(N)$ depends on details of a specific network. 
For example, for the configuration model without multiple connections, 
the dependence $q_{\text{cut}}(N) \sim \sqrt{N}$ is usually used if 
$2<\gamma <3$. 
It is this function that must be substituted into Eqs.~(\ref{kh}), (\ref{Mh}), and (\ref{k-thr}) below.  
A detailed analysis of Eq. (%
\ref{fk}) shows that the cutoff dramatically changes the behavior of the
function $f_{k}(R)$ near $R=1$. $f_{k}(R)$ has a maximum at $R_{\max }\cong
1-(3-\gamma )^{-1/(\gamma -2)}\,k/q_{\text{cut}}$ and tends to zero at $%
R\rightarrow 1$ instead of divergence. As a result, the $k$-core percolation
again becomes to be the hybrid phase transition. The cutoff determines the
highest $k$-core: 
\begin{equation}
k_{h}\cong p(\gamma -2)(3-\gamma )^{(3-\gamma )/(\gamma -2)}q_{\text{cut}%
}(q_{0}/q_{\text{cut}})^{\gamma -2}.  \label{kh}
\end{equation}
The sizes of the $k$-core at $q_{0}\,\ll \,k\,\ll \,k_{h}$ are given by Eq.~(%
\ref{k-core2}). The relative size of the highest $k$-core is 
\begin{equation}
M(k_{h})\cong p[(3-\gamma )^{-(\gamma -1)/(\gamma -2)}\!-\!1](q_{0}/q_{\text{%
cut}})^{\gamma -1}.  \label{Mh}
\end{equation}
Finally, the threshold of the $k$-core percolation is 
\begin{equation}
p_{c}(k)=1/f_{k}(R_{\max })\cong k/k_{h}.  \label{k-thr}
\end{equation}
If $k\!\rightarrow \!k_{h}$, then $p_{c}(k)\rightarrow 1$, i.e. even minor
random damage destroys the highest $k_{h}$-core. By using exact Eqs. (\ref{R}%
) and (\ref{k-core}), we calculated numerically $M(k)$ and $S(k)$ for a
scale-free network with $\gamma =2.5$, see Fig. \ref{kCores}. These curves
agree with asymptotic expressions (\ref{k-core2}) and (\ref{Mh}).

\emph{$k$-core organization of the router-level Internet.}---%
We 
consider 
the router-level Internet which has lower degree-degree correlations
than the Internet at the Autonomous Systems (AS) level. We substitute the
empirical degree distribution of the router-level Internet as seen in
skitter and iffinder measurements \cite{CAIDA} into our exact equations and
compare our results with the direct $k$-core decomposition of this network.
The calculated sizes of $k$-cores and $k$-shells are shown in Fig.~\ref%
{kCores}. The calculated dependence $S(k)$ [Fig.~\ref{kCores}(b), the IR
curve] is surprisingly similar to the dependence obtained by the direct $k$%
-core decomposition of, actually, a different network---the AS-level
Internet---in Ref. \cite{k05}. On the other hand, one can see in Fig.~\ref%
{kCores} that the highest $k$-core with $k_{h}=10$ occupies about 2\% of the
network, while a direct $k$-core decomposition of the same router-level
Internet map in Ref. \cite{aabv05} revealed $k$-cores up to $k_{h}=32$. This
difference indicates the significance of degree--degree correlations, which
we neglected.

\emph{Discussion and conclusions.}---It is important to indicate a quantity critically divergent at the $k$-core's birth point. This is a mean size of a cluster of vertices of the $k$-core with exactly $k$ connections inside of the $k$-core. One may show that it diverges as $-dM(k)/dp \sim (p-p_c)^{-1/2}$ and that the size distribution of these clusters is a power law at the critical point. 

One should note that the $k$-core (or bootstrap) percolation is not related to the recently introduced $k$-clique percolation \cite{dpv05} despite of the seemingly similar terms. The $k$-clique percolation is due to the overlapping of $k$-cliques---full subgraphs of $k$ vertices---by $k-1$ vertices. Therefore, the $k$-clique percolation is impossible in sparse networks with few loops, e.g., in the configuration model and 
in classical random graphs, 
considered here.   

In summary, we 
have developed the theory of $k$-core percolation
in damaged uncorrelated networks. We have found that if the second moment of
the degree distribution of a network is finite,  
the $k$-core transition has the hybrid nature. 
In contrast, 
in 
the networks with
infinite $z_{2}$, instead of the hybrid transition, we have observed 
an 
infinite order transition, similarly to the ordinary percolation in this
situation. All $k$-cores in these networks are extremely robust against
random damage. It indicates the remarkable robustness of the entire $k$-core
architectures of infinite networks with $\gamma \leq 3$. Nonetheless, we
have observed that the finite networks are less robust, and increasing
failures successively destroy $k$-cores starting from the highest one. Our
results can be applied to numerous cooperative models on networks: a
formation of highly connected communities in social networks, the spread of
diseases, and many others.

This work was partially supported by projects POCTI: FAT/46241, MAT/46176,
FIS/61665, and BIA-BCM/62662, and DYSONET. The authors thank D. Krioukov of
CAIDA for information on the Internet maps\vspace{-6pt}.


\begin{thebibliography}{99}
\bibitem{remark} Note that due to the tree-like structure of the infinite sparse configuration model, finite $k$-cores do not exist.
\end{thebibliography}
\end{document}